\documentclass[useAMS,usenatbib]{mn2e}
\usepackage{mathabx}
\usepackage{booktabs}

\title[Helium line emissivities for nebular astrophysics]
{Helium line emissivities for nebular astrophysics}
\author[Del Zanna and Storey]{G. Del Zanna$^{1}$\thanks{E-mail: gd232@cam.ac.uk},
P.~J. Storey$^{2}$\\
$^{1}$ DAMTP, Centre for Mathematical Sciences, University of Cambridge, Wilberforce Road, Cambridge CB3 0WA, UK \\
$^{2}$ Department of Physics and Astronomy, University College London, London WC1E 6BT, UK \\
}
\date{Submitted to MNRAS  }

\pagerange{\pageref{firstpage}--\pageref{lastpage}} \pubyear{2016}

\DeclareMathAlphabet{\mathsc}{OT1}{cmr}{m}{sc}
\def\testbx{bx}%
\DeclareRobustCommand{\ion}[2]{%
\relax\ifmmode
\ifx\testbx\f@series
{\mathbf{#1\,\mathsc{#2}}}\else
{\mathrm{#1\,\mathsc{#2}}}\fi
\else\textup{#1\,{\mdseries\textsc{#2}}}%
\fi}

\usepackage{graphicx,threeparttable}
\usepackage{txfonts}

\usepackage{epsfig}
\usepackage{natbib}
\usepackage{color}

\begin{document}

\label{firstpage}
\maketitle

\begin{abstract}
We present the results of several collisional-radiative models 
describing optically-thin emissivities of the main lines in neutral 
helium formed by recombination, for a grid of electron temperatures and densities, typical of H II regions and Planetary Nebulae. 
Accurate emissivities are required for example to measure the helium 
abundance in nebulae and as a consequence its primordial value.
We compare our results with those obtained by previous models, 
finding significant differences, well above the target accuracy 
of one percent. We discuss in some detail our chosen set of 
atomic rates and the differences with those adopted by 
previous models. The main differences lie in the treatment of electron and proton collision rates and we discuss which transitions are least sensitive to the choice of these rates and therefore best suited to high precision abundance determinations.
We have focused our comparisons on the case B approximation where only He and He$^+$ are considered, but also 
present results of full models including the bare nuclei,
photo-excitation and photo-ionisation 
and either black-body or observed illuminating spectrum in the 
case of the Orion nebula, to indicate which spectral lines are
affected by opacity. For those transitions, accurate
radiative transfer calculations should be performed. 
We provide tables of emissivities for all transitions within $n \le 5$ 
and all those between the $n \le 5$ and $n' \le 25$ states,
in the log $T_{\rm e}$ [K]=10$^{3.0(0.1)4.6}$ and log $N_{\rm e}$ [cm$^{-3}$]=10$^{2(0.5)6}$
ranges, and a FORTRAN code to interpolate to any $T_{\rm e}, N_{\rm e}$ within these ranges. 
\end{abstract}


\section{Introduction}

Helium lines in the visible and near infrared
are particularly important for nebular astrophysics, for
example being routinely used 
to measure the helium abundance in different astrophysical
  sources, and then extrapolating the results to obtain a measurement
  of its primordial abundance
  \citep[see, e.g.][ and references therein]{peimbert_etal:2007,izotov_etal:2014,aver_etal:2015}.
  Such measurements are very important as they provide
  constraints on e.g. Big Bang Nucleosynthesis models
and galactic chemical evolution.

As spectral line intensities can be measured within an uncertainty 
of 1\% or so, a similar accuracy has been sought in atomic modelling.
As we describe below, a significant effort was put in place by
various groups to try to achieve such accuracy, by improving the modelling
and the basic atomic rates.  However, significant differences
(up to 50\% or so) are found in the different calculations.

Recently, we constructed a new model for the level populations of, and resulting line emission from, helium in
the relatively high-temperature, high-density plasma of the solar corona
\citep{delzanna_etal:2020_he}. We reviewed the basic atomic rates 
and found some shortcomings in the rates adopted by previous authors.
In the present paper we extend that model to predict helium
line emissivities in the relatively low-temperature,  low-density photoionized plasma
typical of nebulae, and provide some comparisons with the previous models.

In the next Section we provide a brief overview of some of the most
widely-used previous models. In the following Section we present a
summary of the rates adopted and the various atomic models we have built.
A sample of results and comparisons with previous models is
presented in Section~4, while Section~5 draws the conclusions.

\section{Previous models}

Various models of helium emission in the conditions prevailing in photoionized plasma were developed in the 1950s and 1960s
by several authors, including
\cite{mathis:1957ApJ...125..318M,seaton:1960MNRAS.120..326S,robbins:1970ApJ...160..519R}.
 Most of the early theory and
methods adopted by subsequent authors
are due to Seaton and Burgess, see e.g.
\cite{burgess_seaton:1960MNRAS.120..121B,burgess_seaton:1960MNRAS.121..471B,seaton:1962,pengelly_seaton:1964}.

The details of the various atomic models are not always
entirely clear. However, we now summarise the main assumptions
and rate coefficients adopted by the various authors.

The first detailed model of the He recombination spectrum is that of
\cite{brocklehurst:1972} (B72).
He built a model with the rates available at the time,
and provided emissivities in the case A and B approximations \citep[see][]{baker_menzel:1938}. 
For the $l$-changing collisional excitation (CE) rates, the
semi-classical impact parameter approximation of \cite{seaton:1962} (S62) and \cite{pengelly_seaton:1964} (PS64) was used. 
He used the non-degenerate formulation (S62, hereafter IP) for  $l \leq 5$, and the degenerate formulation (PS64) for higher $l$. 
A few shortcomings in the model were pointed out by subsequent authors,
the main one being that \cite{brocklehurst:1972} neglected
the metastability of the 2$^3$S, 2$^1$S states. 
The approach  was  to solve the statistical balance equations in terms
of departure coefficients $b$ from Saha-Boltzmann level populations
(the so-called  $b$-factors). The calculations treated the singlet and triplet series separately and 
considered first  $n$-resolved levels, $b_n$ and then the $b_{nl}$ for the terms (in  $LS$ coupling)
were calculated for a lower set.

\cite{almog_netzer:1989} (AN89)
built a model with  $LS$ resolved states up to $n=10$ (singlets)
and $n=12$ (triplets), and added  four collapsed levels \citep{burgess_summers:1969} to mimic the
presence of levels up to $n=100$, with the usual assumption that
the collapsed states contain levels that are statistically populated.
As the authors state, this model
is a good assumption for high densities only, which was the main
topic of that paper. Indeed  collisional ionization was added into
the model, as it is important for higher densities.

\cite{smits:1996} (S96) built an $LS$ model,
improving and correcting some errors in his previous models.
Some details of the model can  be found in  \cite{smits:1991a} (S91):
many rates were calculated with the hydrogenic approximation.
The method followed B72 and first calculated the $b$-factors 
for the $n$-resolved states up to $n=496$. 
A matrix condensation technique \citep{burgess_summers:1969}
was  employed to reduce the number of levels to 100. 
The  $b_{nl}$ factors were then calculated for a set of 
$LS$-resolved states up to $n=50$ (2549 levels).
The author then provided line emissivities calculated in case A and B
for  low densities, with the assumption that levels above $n=50$
are statistically redistributed.
As the author pointed out, this assumption is 
 {\it not always valid}.  The $b_{nl}$ at $n=50$ are not
 equal to the $b_n$ for e.g. a density of 100 cm$^{-3}$.
 Smits argued that  errors introduced by this assumption are not large,
but without actually quantifying the statement.  
We will return to this issue below, as we have built several
models and are able to assess this assumption.

S96 used for the $n \le 9$ states the A-values obtained
from the oscillator strengths calculated by
\cite{kono_hattori:1984}. For higher states, the Coulomb
approximation was used for $l \le 2$ states,
and for the others a scaled hydrogenic approximation
was adopted.
CI rates were not included.
The radiative recombination (RR) rate
coefficients for lower $n$ states were obtained from the OP photoionization
(PI) cross-sections of \cite{fernley_etal:1987}.
For higher $n$,  scaled hydrogenic rates were used.
S91 states that  CE rates within the  $n=2,3$ levels are taken
from the $R$-matrix calculations of \cite{berrington_kingston:1987} (BK87).
They were used to find the populations
of the $n=2,3$ levels, to include the metastability of the 2s $^3$S, which was not
included by   Brocklehurst (1972).
The rest of the CE rates were taken from Brocklehurst (1972), i.e. using the impact parameter
approximation.
{\cite{smits:1996} noted that transfer of population between singlet
and triplet states by electron collisions would be included if $R$-matrix
CE rates for the lower states were used. This mixing was
included in his model only for the $n=2$ states. 
Spin-orbit and other relativistic interactions between $^3$L$_{L}$ and $^1$L$_{L}$ states for high $L$ are also
real effects not taken into account by \cite{smits:1996}.} Both effects could
reduce the populations of the triplets, compared to
the singlets. However, \cite{smits:1996} noted that comparisons
with observations indicated an opposite trend, i.e.
some of the predicted intensities of the triplets were lower
than observations.

{ Drawing on the high precision intermediate coupling variational calculations of helium structure and radiative processes by \cite{drake:1996},  \cite{bauman_etal:2005} built an $LSJ$ model of helium
in the low-density limit (i.e. no collisions), and
concluded  that singlet-triplet mixing has a negligible effect on the total intensities
of the lines within a multiplet. However, it is still unclear
what effects collisional processes linking singlets and triplets have.
As discussed below, we include these processes in our model up to $n=5$ states.}
\cite{bauman_etal:2005} also pointed out that
individual intensities within a multiplet
would be affected, but are generally not observable as line
widths are larger than the separation of lines.  
A simpler $LS$ model is therefore equivalent to an $LSJ$ model.

\cite{benjamin_etal:1999} (B99) built a  case B
$LS$ model with  only the lower 29 states (up to $n=5$), using the
\cite{sawey_berrington:1993} 
(SB93) $R$-matrix cross-sections for up to $n=4$, with some
interpolations and extrapolations.
For the $n=5$ states, IP rates for $l \le 2$ and hydrogenic
values for higher $l$ were used. 
The A-values were mostly the same as in the S96 model,
as well as the RR rates.
For $\Delta n =0$ and $n \ge 2$ CE rates for electrons, protons and
He$^+$ were included, using the IP method for low $l$ and the
PS64 otherwise, assuming equal proton and electron densities $N$, and
N(He$^+$)=0.1 $N$.
For $\Delta n =0$ and $n=2$ the IP rates for collisions with
protons and He$^+$ were included. 
For the CI, B99 included a rate for the 2$^3$S metastable level
and the \cite{vriens_smeets:1980} (VS80) semi-classical estimates for
the other levels.
The model built by B99 is much  {reduced in size } compared to that presented by S96.
To improve it,  'cascading'
(described as an indirect recombination) contribution
from the higher states was estimated so as to match the
S96 populations within 2\% for densities lower than 10$^6$.
However, as we show below, much larger differences between B99 and s96
are actually present in the line emissivities.

\cite{porter_etal:2005} (P05) constructed an $LS$ model up to $n^*$,
with an extra `collapsed' $n^* +1$ level describing the missing states between
$n^*$ and the continuum, and presented emissivities in the case B approximation.
They noted that with $n^*=100$, the corrections due to the collapsed level
are negligible for low densities.  
Transition probabilities from the nearly exact
calculations of \cite{drake:1996} (up to $n=10$) were used.
PI cross-sections from \cite{hummer_storey:1998} (HS98) were adopted to calculate the RR rates,
while the CE rates of \cite{bray_etal:2000} were used.
For the $l$-changing collisions, they used the IP method of S62
for the s,p,d states, and the semi-classical theory of
\cite{vf:2001} (VF01) for higher $l$. 

\cite{porter_etal:2007} (P07) built a
similar $LS$ model up to $n=40$, and added collapsed $n$-resolved
levels between  $n=41$ and $n=100$, i.e. similar to our coronal
model for helium. They applied this model to study the helium
abundance in the Orion nebula.

\cite{porter_etal:2012} (P12) presented an updated case B model,
with the full set of HS98 PI cross sections (up to $n=25$) and associated
RR rates. For higher $n$ states, hydrogenic RR rates were used.
The model included $LS$ states up to $n=100$, and a single collapsed 
$n=101$ level. The paper reports that a code error in their earlier calculations (P05,P07)
was uncovered, which affected mostly the 5876, 6678~\AA\ lines, the decays from the singlet
and triplet 3d levels.
Fixing the code error mostly increased the recombination coefficients into the 3d levels.
Other differences were due to a different implementation of the semi-classical
\cite{vriens_smeets:1980} collisional excitation rates. 

The P05, P07, P12 models were built within the various improved versions
of the He model within the  CLOUDY \citep{ferland_etal:2017} photoionization code.
{The semi-classical theory of VF01 has been discussed by \cite{guzman_etal:2017} who show that due to the truncation of the cross-sections at energies that neglect the quantum mechanical tail, it grossly underestimates the collision rates, by a factor of six at $n=30$, for example. \cite{guzman_etal:2017} also show that the  PS64 method, on the other hand, gives rates close to those obtained from a quantum mechanical treatment. 
 }   

Another point worth noting is that the IP proton rates as shown in
\cite{guzman_etal:2017} and calculated within CLOUDY are also
incorrect, as we noted in our coronal model paper \citep{delzanna_etal:2020_he}.
{We also found that the Bray et al. CE rate file in CLOUDY inverted by mistake the values for the transitions from the metastable 2s $^3$S to the 4s $^3$S and 4p $^3$P (levels 12,14), thus affecting somewhat the emissivities of the main decays from these two levels, the 3188 and 4713~\AA\ lines. }
{Thus all the previous helium recombination models have apparent defects, in terms of using rates coefficients that are now considered not accurate or the best available at present. The purpose of this paper is therefore to investigate whether correcting these shortcomings has any impact on the emissivities of the spectroscopically important transitions.}   

\section{Models}

The present models are an extension of our previous  coronal models
described in \cite{delzanna_etal:2020_he}.
Among them, the most extended neutral He model for low-temperature ($T$=20000 K) 
plasma  was a set of $LS$-resolved states up to $n=40$,
and a set of $n$-resolved levels up to $n=100$.
Considering the behaviour of the $b$ factors, this model was deemed
sufficient for electron densities higher than  $N$=10$^{6}$ cm$^{-3}$. 

For the present paper we have built a model for
neutral He with $LS$-resolved states up to $n=100$ and $n$-resolved levels up to $n=500$.
This is a much larger model than the previous ones, and especially larger than the
largest model ($LS$-resolved states up to $n=50$) produced by S91 and
subsequently used by S96 and B99. 
For He$^{+}$ we adopted the $J$-resolved CHIANTI \citep{dere_etal:2019}  model.
We create a collisional-radiative model, with
matrices that contain  all the main rate coefficients affecting the
bound levels, and obtain the level populations in equilibrium by direct inversion.

We did not include dielectronic recombination (DR), the key process in
our  coronal model, 
as it is negligible at the low temperatures of interest here. {For RR and $L\leq 3$, we use rates obtained by numerical integration of the photo-ionization cross-sections calculated by \cite{hummer_storey:1998} in the R-matrix approximation with a target that accounted for the dipole and quadrupole polarisabilities of the He$^+$ ground state. \cite{hummer_storey:1998} showed that their calculated bound-free cross-sections agree within 1\% with the bound-bound radiative data of \cite{drake:1996} when extrapolated to the series limit. Hydrogenic values were used for RR for $L > 3$.}

For the $LS$ levels up to $n=10$ we have used the  energies and A-values
of \cite{drake_morton:2007}. The energies for the higher levels were obtained
from the quantum defects using the updated coefficients of \cite{drake:2006}. The A-values for the higher $LS$ states were obtained either from fits to the results of  \cite{drake_morton:2007}, as described by \cite{hummer_storey:1998}, or using the methods of 
 \cite{bates_damgaard:1949}, \cite{vanRegemorter:1979}, and the hydrogenic approximation using the code RADZ1 \citep{storey_hummer:1991}.
 The A-values for the $n$-resolved states to the lower $n=2,3$ states were 
 obtained from statistically weighted averages of the extrapolated \cite{drake_morton:2007}. The A-values for the $n$-resolved states to the lower $n=4,40$ states were obtained by averaging hydrogenic values. 
 The A-values between the $n$-resolved states were obtained as in our 
 previous coronal model, using the hydrogenic analytical formulation and 
 tabulated Gaunt factors.

 For the electron CE rates of states up to $n=5$
 we use the $R$-matrix results of \cite{bray_etal:2000},
 although we have also experimented with other rates.
 The lowest temperature for the \cite{bray_etal:2000} rates is 5623 K. 
 To provide estimates to lower temperatures, we have proceeded as follows. 
 For the CE rates within the n=2,3 states, we have taken the $R$-matrix rates from BK87, which were calculated down to 1000 K. As some small differences between the two sets of rates  are present, we have interpolated the BK87 rates to 5623 K, and scaled them so they agree at this temperature with the 
 Bray et al. values. For the n=4 states, we have scaled the SB93 rates in a similar way, and then
 extrapolated them down to 1000 K in the \cite{burgess_tully:1992}
 scaled domain, with a linear fit 
 to the first two points (SB93 CE rates were calculated for 2000 and 5000 K). 
 For the n=5 states, we have linearly extrapolated the Bray et al. CE rates in the 
  Burgess and Tully scaled domain, considering again the first two points. 
  The CE rates were stored 
in the  scaled domain, where the interpolation in temperature
is carried out, as done within the CHIANTI software.

  Note that the collision strength calculations of \cite{sawey_berrington:1993} and of \cite{bray_etal:2000} both employ the R-matrix method but the latter calculation is to be preferred because it makes allowance for the presence of continuum target states with the result that their collision rates between bound states are almost always smaller than those of \cite{sawey_berrington:1993} due to flux lost to the continuum states. It should be noted, however, that the allowance for continuum states gives rise to some resonance features near threshold that, although they represent a physical process, are not necessarily correct in detail. Thus a measure of uncertainty is still attached to the CE between low-lying states that is difficult to quantify. 
  Finally, we point out that extrapolating the rates to temperatures lower than 1000 K is feasible   but the values would be quite uncertain. 
 
 {To connect the states with $n \leq 5$ to higher states, $ m > 5$, we extrapolated  the cross-sections of  \cite{bray_etal:2000} for $n < m \leq 5$ assuming they vary as $m^{-3}$. } Attempts to directly calculate CE rates to the higher
 levels proved unreliable, as discussed in \cite{delzanna_etal:2020_he}.
 For the other states we used the IP method for $\Delta n=1$ transitions. 
 For $\Delta n=2,3,4$ transitions within the the $n$-resolved levels, we adopted the
 the  \cite{percival_richards:1978} approximation.

 For the $l$-changing ($\Delta n=0, \Delta l=1$)  collisions for electrons and protons
 we used the the IP approximation among the non-degenerate levels
 with lower $l$  and the \cite{pengelly_seaton:1964} (PS64) for the remainder.
{The switching was applied when the difference in energy reached
10$^{-4}$ cm$^{-1}$. We note that 
an improved PS64 method has been developed by Badnell (2021, in press). We have checked that differences with the PS64 rates
are negligible. 
We assumed a fixed proton to electron ratio of 0.91,
which results from assuming that the helium abundance 
is 10\% by number that of hydrogen.

 We included collisional ionization (CI) and three-body recombination
 as described in our previous model for the lowest states, but use 
 the semiclassical CI rates of \cite{vriens_smeets:1980} instead of 
 those developed by A. Burgess (ECIP). We recalculated the 
 rates for the low temperatures considered here.
 We note that CI has a negligible
 effect at the low densities considered here,
 as the model is driven by photoionization (PI) and resulting RR cascading.
 }
 For the PI cross-sections we have used the 
 \cite{hummer_storey:1998} results, instead of the 
 semi-classical Kramers
 hydrogenic formula and the Gaunt factors from \cite{karzas_latter:1961}
as  in our previous paper.
The key factors are the RR rates, the CE rates within the lower
levels (for higher densities), the set of A-values,  and  the cascading
effects from the higher to the lower states.
 Further details on other specific
 rates are given in \cite{delzanna_etal:2020_he}.
 
The emissivities are defined as 
\begin{equation}
E = \frac{4\, \pi \, j}{ N_{\rm e} \, N(He^+) } =  
\frac{h\, \nu_{ji} \, N_j \, A_{ji} \,  N(He)}
{N_{\rm e} \, N(He^+) } 
\quad {\rm erg \, cm}^3 \, {\rm s}^{-1}
\end{equation}
\noindent 
where the first definition is how the emissivities are usually
indicated in the literature, and the second one is how we calculated them: 
$N_j$ is the  population of the upper level $j$, relative to the total population of
He, N(He), i.e. He$^{0}$ and He$^{+}$; $A_{ji}$ is the radiative transition probability,
and $h\, \nu_{ji}$ is the energy of the photon.

We calculate the emissivities in case A (optically thin plasma)
and case B,  which is an approximation to model the real plasma emission when
the strongest singlet lines, decaying to the ground state, are re-absorbed on the spot.
To obtain case B, we have set to zero the RR to the ground state and
all the A-values of the singlet (above $n$=2) decays to the ground state.

{
Clearly, detailed modelling allowing for radiative transfer of line and continuum photons is needed to study
specific sources. Here, we are mostly interested in showing how
line emissivities are affected by the use of different models and atomic rates. 
}

Our general model includes all three ionization stages of He but to make a direct comparison to earlier tabulated emissivities,
we make the simplifying assumption adopted in previous models, where
only neutral He and He$^{+}$ (and not the bare nuclei) are considered, to mimic the conditions in the He$^+$ zone of a photoionized nebula.
Various options are available to achieve this,
forcing the relative numbers of neutral He and He$^+$, for example  using a `laser' monoenergetic photon source (G. Ferland, priv. comm.). 
In a similar way, we construct the present Case B model (PB in the emissivity tables), in which  we include a 
dilute black-body radiation field with  $T$=100,000 K which is used to  photoionize only the ground state. The function of this field is to establish a physically reasonable He$^+$ fraction and a range of choices of temperature and dilution factor are possible that have this effect.  Parameters were chosen that approximate the conditions in a planetary nebula surrounding a hot white dwarf.  We also removed the photons below 228~\AA\ which would normally be absorbed in ionizing He$^+$ closer to the ionizing star and not present in the zone where the He recombination spectrum is formed.
In this model we do not include photo-excitation (PE) effects,
as also assumed  in earlier work on the Case B helium recombination spectrum. 
In this way, the emissivities are independent of the chosen
dilution factor.

To assess how reliable a reduced model (similar to that of S96) is for nebular densities, 
we had initially considered the earlier coronal model, with $LS$-resolved states up to $n=40$
and n-states up to n=100. However, for the lowest densities considered here
(10$^{2}$ cm$^{-3}$), we found that 
the $b$-factors reach unity only around n=200, so we have therefore built 
a new model, which we call PB40, with $LS$-resolved states up to $n=40$ 
and n-states up to n=300. The rates for this model are essentially the same as those of the full 
$n=500$ model. We used the same black-body flux to photoionize the 
ground state to establish a reasonable ionization balance.
Also, to show the effects that different CE rates can have, we have also modified the PB40 model by replacing the Bray et al. CE rates with those 
 from \cite{sawey_berrington:1993}
for states up to $n=4$, and IP rates for the
$n=5$ levels, to approximate the rates used by B99 (although some differences
are present). We call this model SB93. 

Finally, we have also built another case B model (which we call PB+PE+PI)
where we have considered the level balance between 
He, He$^+$ and He$^{++}$ still using the same  black-body flux,
and including both PI and PE. Photoexcitation 
(and de-excitation) was included for all allowed transitions.
PI was included for all states up to $n$=25 using the PI cross sections 
from \cite{hummer_storey:1998}, although we note that only those
for the ground state and the metastable are relevant here.
These cross-sections are close (within a few percent) of those 
in TOPBASE. The photon energies have been adjusted to the observed thresholds.
As shown below, for sufficiently low  dilutions, the results of this 
model are close to those of the PB model.

\section{Results}

\begin{figure}
\centerline{\epsfig{file=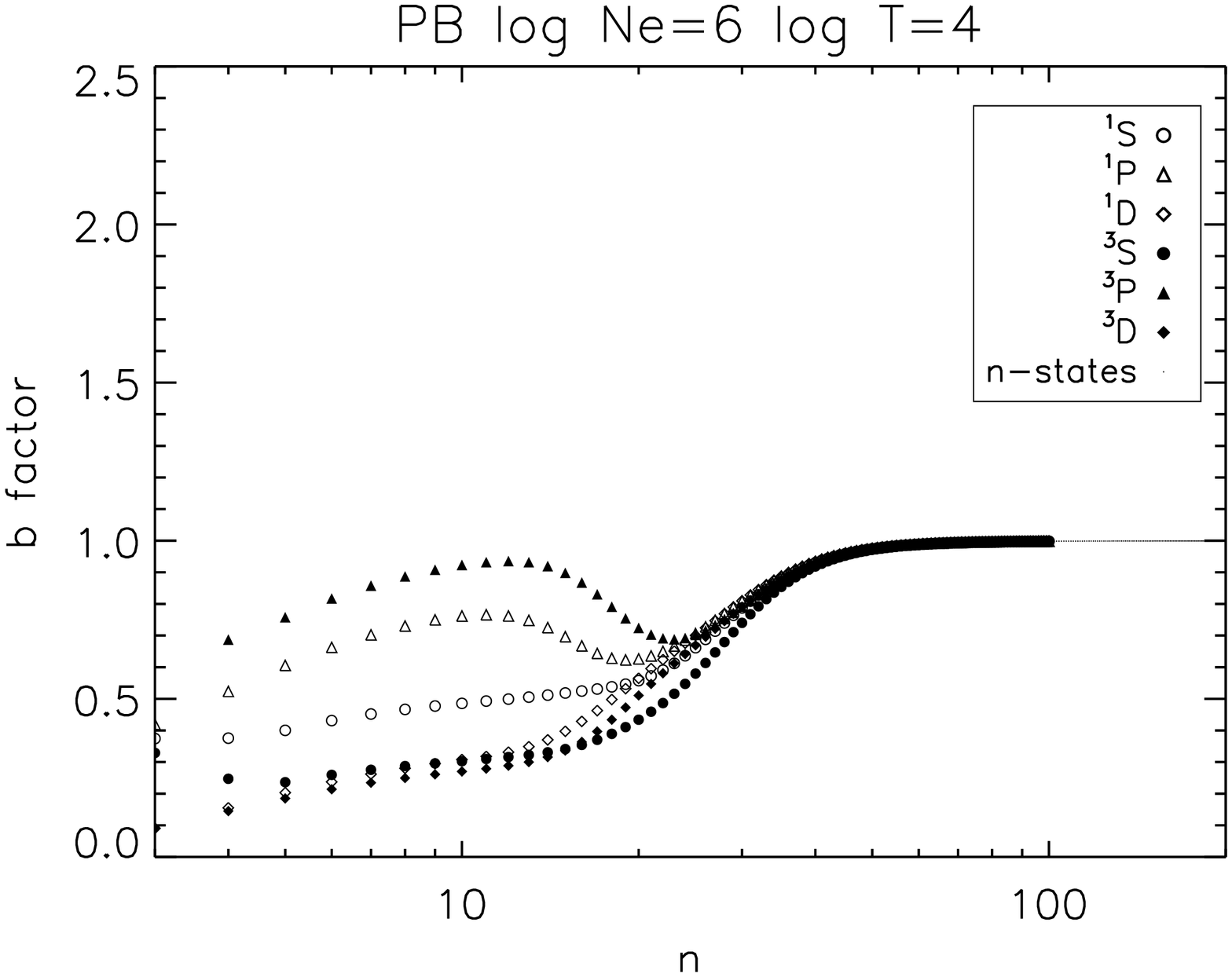, width=7.cm,angle=-90 }}
\centerline{\epsfig{file=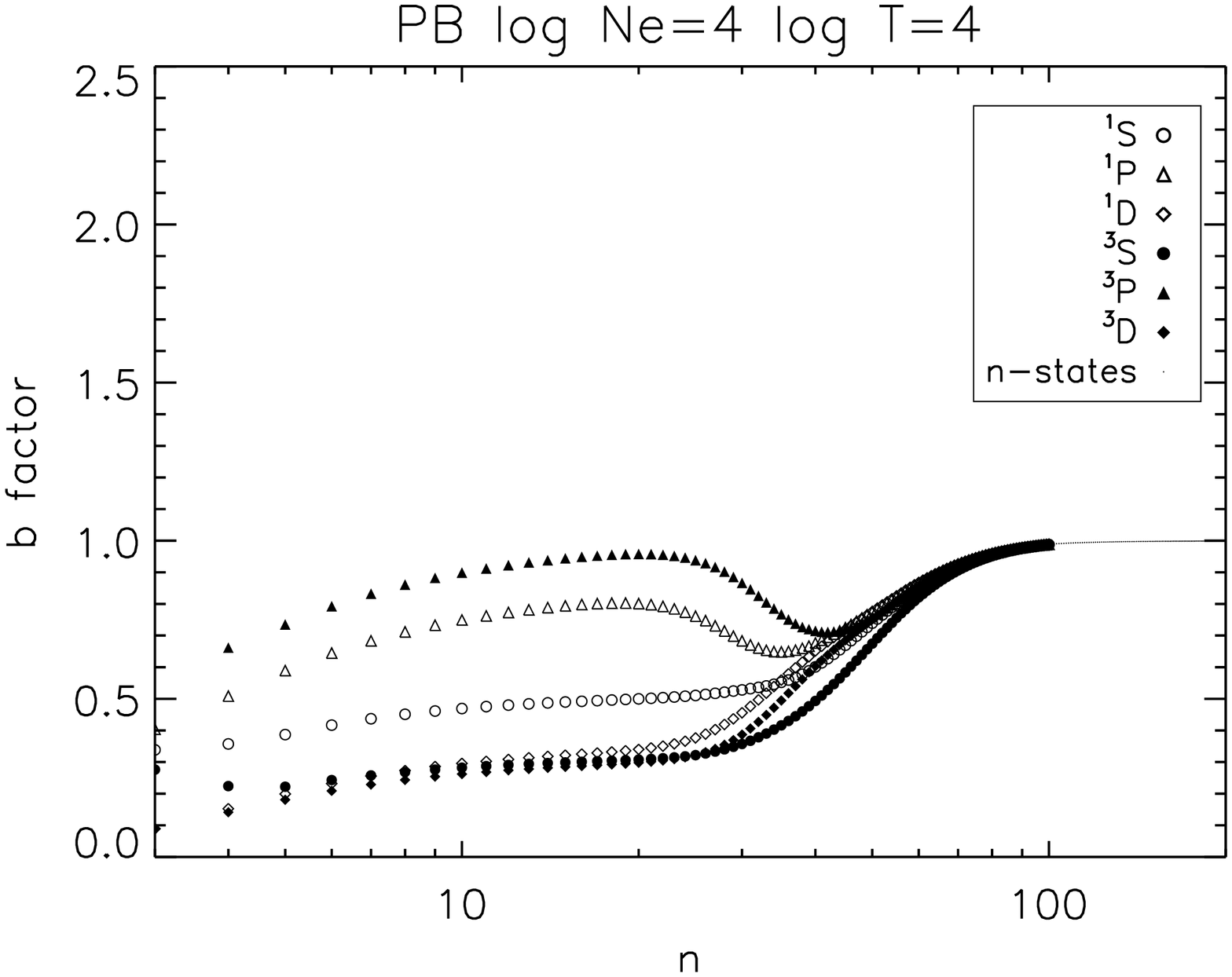, width=7.cm,angle=-90 }}
\centerline{\epsfig{file=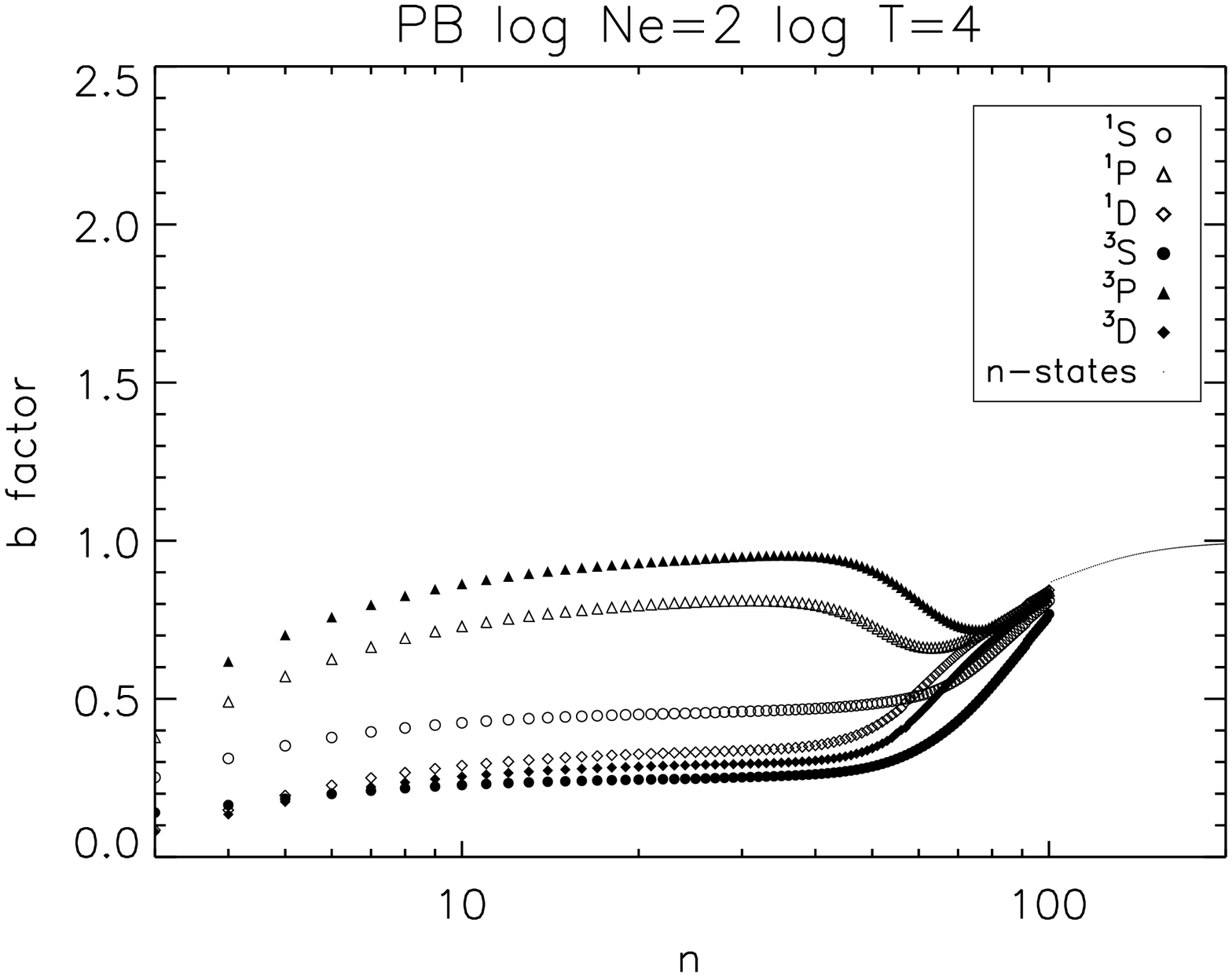, width=7.cm,angle=-90 }}
\caption{$b$ factors obtained with our full case B model,
  $LS$-resolved states up to $n=100$ and $n$-resolved levels up to $n=500$. Only values up to
$n=200$ are shown.}
\label{fig:b1}
\end{figure}

\begin{table*}
\begin{center}
   \caption{Emissivities  (10$^{-26}$ erg cm$^{3}$ s$^{-1}$) of the
     strongest He lines, for $T_{\rm e}$=10000 K and  $N_{\rm e}$=10$^{2}$ cm$^{-3}$
     in case B. 
\label{tab:t1e4_n1e2} }
\begin{tabular}{rllllllllll}
  \hline
 $\lambda$ (\AA) & levels & \multicolumn{5}{c}{Earlier work} & \multicolumn{4}{c}{Present work} \\ 
 \cmidrule(lr){3-7}\cmidrule(lr){8-11}
       &   & B72   & S96 & B99    &P05   & P12 & SB93  & PB40 & PB &  PB+PI+PE \\
  \hline
2945 & T 5p-2s & 2.66&  -    & 2.70  & 2.69  & 2.70 & 2.71 & 2.71 & 2.70 & 2.70 \\
3188 & T 4p-2s & 5.55& 5.61  & 5.62  & 5.62  & 5.62 & 5.62 & 5.62 & 5.61 & 5.61 \\

3889 & T 3p-2s & 13.7& 13.6 & 13.7  & 14.1  & 14.0 & 14.0 & 14.0 & 14.0 & 14.0 \\ 
3965 & S 4p-2s & -   &1.39  & 1.39  & 1.41  & 1.42 & 1.42 & 1.42 & 1.41 & 1.41 \\ 

4026 & T 5d-2p & 2.87& 2.86  & 2.86  & 2.92  & 2.91 & 2.91 & 2.91 & 2.91 & 2.91 \\ 
4388 & S 5d-2p & -   & 0.76  & 0.76  & 0.77  & 0.77 & 0.77 & 0.77 & 0.77 & 0.77 \\ 
4471 & T 4d-2p & 6.05& 6.15  & 6.16  & 6.12  & 6.12 & 6.13 & 6.12 & 6.12 & 6.13 \\ 

4713 & T 4s-2p & 0.56& 0.63  & 0.65  & 0.65  & 0.64 & 0.65 & 0.65 & 0.65 & 0.65 \\ 
4922 & S 4d-2p & -   & 1.63  & 1.64  & 1.66  & 1.65 & 1.65 & 1.65 & 1.63 & 1.65 \\ 

5016 & S 3p-2s & -   &3.48  & 3.49  & 3.54  & 3.56 & 3.55 & 3.55 & 3.55 & 3.55 \\ 
5876 & T 3d-2p & 16.7& 16.8   & 18.90 & 16.3  & 16.9 & 16.9 & 16.9 & 17.0 & 17.0 \\ 
6678 & S 3d-2p & -   & 4.78 & 4.79  & 4.63  & 4.80 & 4.80 & 4.80 & 4.83 & 4.83 \\ 

7065 & T 3s-2p & 2.0& 2.83 & 2.96  & 3.0   & 2.98 & 2.99 & 2.99 & 2.98 & 2.98 \\ 
7281 & S 3s-2p & -  & 0.89 & 0.90  & 0.89  & 0.90 & 0.90 & 0.90 & 0.90 & 0.90 \\ 
10830 & T 2p-2s& 26.8&  34.3 & 34.0  & 33.2  & 33.4 & 33.8 & 33.5 & 33.5 & 33.6 \\ 

18685 &T 4f-3d & -   &  -   & 2.22  & 2.18  & 2.20 & 2.21 & 2.21 & 2.24 & 2.23 \\ 
20587 &S 2p-2s & -   & 4.12 & 4.13  & -     & 4.15 & 4.17 & 4.16 & 4.17 & 4.17 \\ 
\hline
\end{tabular}
\begin{tablenotes}
    \item[] {The first column gives the wavelength
  (in air, except the  last ones in vacuum), the second indicates if a line is
      between singlets (S) or triplets (T).
      B72: Brocklehurst (1972);
      S96:   Smits (1996);  
      B99: Benjamin et al. (1999);         
         P05: Porter et al. (2005);
         P12: Porter et al. (2012);
 SB93: $n$=40 model with  the SB93 rates; 
          PB40: $n$=40;
          PB: full n=500;
          PB+PI+PE: full n=500 model with PI and PE and dilution d=10$^{-16}$.
}
 \end{tablenotes}

\end{center}
\end{table*}

\begin{table*}
\begin{center}
   \caption{Emissivities  (10$^{-26}$ erg cm$^{3}$ s$^{-1}$) of the
     strongest He lines, for $T_{\rm e}$=10000 K and  $N_{\rm e}$=10$^{4}$ cm$^{-3}$
     in case B.
\label{tab:t1e4_n1e4} }
\begin{tabular}{@{}rlllllllllllllllllllll@{}}
  \hline
 $\lambda$ (\AA) & levels & \multicolumn{5}{c}{Earlier work} & \multicolumn{4}{c}{Present work} \\ 
 \cmidrule(lr){3-7}\cmidrule(lr){8-11}
& &
                   B72&  S96 & B99    &P05    & P12  & SB93 & PB40 & PB &  PB+PI+PE\\
  \hline
2945 & T 5p-2s & 2.67 &  -    & 2.73   &  2.82 & 2.84 & 2.83 & 2.83 & 2.83 &  2.83 \\
3188 & T 4p-2s & 5.57 & 5.63  & 6.07   &  6.12 & 6.12 & 6.13 & 6.01 & 6.01 &  6.01 \\

3889 & T 3p-2s & 13.7 & 13.7  & 16.7   & 16.8  & 16.6 & 17.0 & 16.7 & 16.7 &  16.7 \\
3965 & S 4p-2s & 1.42 & 1.39  & 1.43   & 1.47  & 1.51 & 1.47 & 1.47 & 1.47 &  1.47 \\

4026 & T 5d-2p & 2.88 &  2.87 & 2.87   &  3.03 & 3.02 & 3.02 & 3.02 & 3.02 &  3.02 \\
4388 & S 5d-2p & 0.78 & 0.76  & 0.76   &  0.80 & 0.81 & 0.79 & 0.79 & 0.79 &  0.79 \\
4471 & T 4d-2p & 6.07 & 6.16  & 6.52   &  6.47 & 6.46 & 6.57 & 6.43 & 6.44 &  6.44 \\

4713 & T 4s-2p & 0.56 & 0.63  & 0.95   &  0.83 & 0.82 & 0.95 & 0.89 & 0.88 &  0.88 \\
4922 & S 4d-2p & 1.66 & 1.64  & 1.69   &  1.72 & 1.77 & 1.71 & 1.70 & 1.70 &  1.70 \\

5016 & S 3p-2s & 3.57 & 3.49  & 3.73   &  3.81 & 3.90 & 3.81 & 3.78 & 3.79 &  3.79 \\
5876 & T 3d-2p & 16.8 & 16.7  & 19.1   &  18.7 & 19.4 & 19.3 & 19.1 & 19.1 &  19.1 \\
6678 & S 3d-2p & 4.80 & 4.75  & 5.05   &  4.96 & 5.45 & 5.10 & 5.06 & 5.06 &  5.06 \\

7065 & T 3s-2p & 1.99 & 2.84  & 6.15   &  5.90 & 5.82 & 6.06 & 5.88 & 5.87 &  5.89 \\
7281 & S 3s-2p & 0.84 & 0.46  & 1.28   &  1.20 & 1.23 & 1.27 & 1.21 & 1.21 &  1.21 \\
10830 & T 2p-2s& 26.8 & 234.4 & 204.7  &  188  & 185  & 196  & 188  & 188  &  189 \\

18685 &T 4f-3d & -   & -      & 2.26   &  2.28 & 2.36 & 2.29 & 2.27 & 2.27 &  2.27 \\
20587 &S 2p-2s & -   & 6.91   & 6.75   &   -   & 6.73 & 6.70 & 6.60 & 6.60 &  6.61 \\
\hline
\end{tabular}
\end{center}
\end{table*}

\

\begin{table*}
\begin{center}
   \caption{Emissivities  (10$^{-26}$ erg cm$^{3}$ s$^{-1}$) of the
     strongest He lines, for $T_{\rm e}$=10000 K and  $N_{\rm e}$=10$^{6}$ cm$^{-3}$
     in case B.
\label{tab:table_t1e4_n1e6} }
\begin{tabular}{@{}rllllllllllllllllllll@{}}
  \hline
 $\lambda$ (\AA) & levels & \multicolumn{5}{c}{Earlier work} & \multicolumn{4}{c}{Present work} \\ 
 \cmidrule(lr){3-7}\cmidrule(lr){8-11}
 &                 &  B72   &S96 & B99   &P05   & P12   & SB93 &  PB40 & PB &  PB+PI+PE & \\
  \hline
2945 & T 5p-2s & 2.72& -     & 2.79 & 2.96 & 2.97  & 2.92 & 2.92 & 2.92 & 2.92 & \\
3188 & T 4p-2s & 5.66& 5.72  & 6.32 & 6.51 & 6.48  & 6.40 & 6.24 & 6.24 & 6.24 & \\

3889 & T 3p-2s & 14.0& 13.9  & 17.9  & 18.3 & 18.1 & 18.3 & 17.9 & 17.9 & 17.9 & \\ 
3965 & S 4p-2s & 1.45& 1.42  & 1.47  & 1.54 & 1.59 & 1.51 & 1.51 & 1.51 & 1.51 & \\ 

4026 & T 5d-2p & 2.93& 2.89  & 2.89  & 3.18 & 3.14 & 3.09 & 3.09 & 3.09 & 3.09 & \\ 
4388 & S 5d-2p & 0.79& 0.77  & 0.76  & 0.83 & 0.85 & 0.81 & 0.81 & 0.81 & 0.81 & \\ 
4471 & T 4d-2p & 6.15& 6.20  & 6.70  & 6.81 & 6.76 & 6.78 & 6.60 & 6.60 & 6.60 & \\ 

4713 & T 4s-2p & 0.56& 0.63  & 1.07  & 0.91 & 0.90 & 1.07 & 0.98 & 0.98 & 0.98 & \\ 
4922 & S 4d-2p & 1.68& 1.65  & 1.73  & 1.80 & 1.85 & 1.75 & 1.73 & 1.73 & 1.73 & \\ 

5016 & S 3p-2s & 3.63& 3.54  & 3.87  & 4.04 & 4.13 &  3.96 & 3.93 & 3.93 & 3.93 & \\ 
5876 & T 3d-2p & 16.8& 16.6  & 19.9  & 20.2 & 20.85& 20.1 & 19.9 & 19.9 & 19.9 & \\ 
6678 & S 3d-2p & 4.80& 4.73  & 5.15  & 5.22 & 5.77 & 5.21 & 5.14 & 5.14 & 5.14 & \\ 

7065 & T 3s-2p & 2.0&  2.86  & 7.34  & 7.17 & 7.00 & 7.25 & 7.00 & 7.00 & 7.00 & \\ 
7281 & S 3s-2p & 0.85& 0.89  & 1.42  & 1.36 & 1.38 & 1.42 & 1.34 & 1.34 & 1.34 & \\ 
10830 & T 2p-2s& 27.1& 320 & 267 & 255 & 247&  259 & 247  & 247  & 247 & \\ 

18685 &T 4f-3d &   -  &   -   & 2.24  & 2.37 & 2.47 & 2.26 & 2.24 & 2.24 & 2.24 & \\ 
20587 &S 2p-2s &   -  & 8.54  & 8.15  & -    & 8.25 & 8.09 & 7.97 & 7.97 & 7.97 & \\ 
\hline
\end{tabular}
\end{center}
\end{table*}

\begin{table}
\begin{center}
   \caption{Emissivities  (10$^{-26}$ erg cm$^{3}$ s$^{-1}$) of the
     $\lambda 10830$ line as a function of temperature ($T_{\rm e}$)  electron density ($N_{\rm e}$) in case B. 
\label{tab:table_10830} }
\begin{tabular}{@{}rllll@{}}
  \hline
$T$[K] & Calc. & \multicolumn{3}{c}{$N_e$ [cm$^{-3}$]} \\
        &   &  $10^2$   & $10^4$   & $10^6$   \\
  \hline
5000 & PB & 50.6    & 130 & 189 \\

     & P12 & 50.8 & 140  & 213 \\

     & P05 & 49.9 & 140  & 214 \\
     
     & B99 & 50.7 & 152  & 234  \\ 
    
10000 & PB & 33.5 & 188 & 247 \\

     & P12 & 33.4 & 185  & 247 \\

     & P05 & 33.2 & 188  & 255 \\
     
     & B99 & 34.0 & 205  & 267  \\
     
     & PB40 & 33.5 & 188 & 247 \\
     
     & SB93 & 33.8 & 196 & 259 \\

20000 & PB & 23.6    & 204 & 256 \\

     & P12 & 23.8 & 181  & 215 \\

     & P05 & 23.4 & 188  & 237 \\
     
     & B99 & 24.6 & 207  & 253  \\ 

\hline
\end{tabular}
\end{center}
\end{table}

We present a sample of results, for a temperature
$T$= 10000 K  and three electron densities,
$N$=10$^{6}$, 10$^{4}$, 10$^{2}$ cm$^{-3}$. 
Figure~\ref{fig:b1} shows the $b$ factors obtained from the
model PB, $LS$-resolved states up to $n=100$ and $n$-resolved levels up to $n=500$
in case B.
$b$ factors only up to $n=200$ are shown, as they have already reached near
unity for $n=100$.  For example, at 10$^{2}$ cm$^{-3}$, the $b$
factors are 0.9968 at $n$=200 and  0.998 at $n$=300.
The figures indicate a smooth behaviour in transitioning from the
$LS$-resolved to the $n$-resolved states. The assumption that
the $n$-resolved  states are in statistical equilibrium is validated.

The resulting emissivities for all the strongest He lines
in the visible/near infrared are shown in the last two columns
of the emissivity Tables 1, 2 and 3  below. Additional Tables are provided 
in the Appendix.
Our final results are in the penultimate column, designated PB. The final column displaying the effect of adding photoionization and photoexcitation of excited states will be discussed later.  

The emissivities with the reduced
$n=40$ model (PB40) are also given in the Tables.  
There is generally excellent agreement with the PB results, with 
the largest difference of 1.7\% for the lowest temperatures and densities.
 This largely confirms the suggestion by S91 that their model
 (which we recall was similar to our PB40 calculation in that it included $LS$-resolved states up to $n=50$) would provide reasonably accurate emissivities even at 10$^2$~cm$^{-3}$ for the most prominent optical lines.

Earlier results are shown in previous columns.
Despite the simplified rates used in the first complete model built by
\cite{brocklehurst:1972}, and the neglect of the metastability of the 2$^3$S, 2$^1$S
states, relatively good agreement for many lines can be seen.
We expected close agreement with the other models,
but that is not the case. There are also surprising large 
differences for some lines between the S96 and B99 models,
contrary to the 
 B99 statement that agreement was present at the 2\% level.

We experimented with changing various rates, and 
found that at these plasma $T_{\rm e}, N_{\rm e}$ the electron CE rates among the lower levels
have a significant effect, as already pointed out in previous literature.
The values of the CE rates are relatively uncertain compared to radiative rates, so we begin by comparing the different calculations at the lowest density, where CE is much less significant. At 10000K, we find average absolute differences of 1.6\% (maximum 5.0\%) compared to S96, 1.5\% (maximum 11.1\%) compared to B99, 1.1\% (maximum 4.1\%) compared to P05 and 0.45\% (maximum 1.8\%) compared to P12. All four of these calculations used highly accurate bound-bound radiative transition probabilities, although from different authors, and differ primarily in their treatment of radiative recombination. As in the current work, the results of P05 and P12 derive radiative recombination rates from the photoionization cross-sections of \cite{hummer_storey:1998} which are to be preferred to the Opacity Project cross-sections used by S96 and B99.     

At typical nebular temperatures, collisional excitation from the ground state is negligible compared to recombination but excitation from the $2~^3$S metastable is significant as temperature and/or density increases. It is instructive to consider the emissivity of the $2~^3$P - $2~^3$S $\lambda 10830$ since it is excited by recombination and cascading through the triplet terms and also strongly affected by collisional excitation from the metastable. Similar but smaller effects of CE are seen for other low-lying triplets and to a much lesser extent, the singlet states. The $\lambda 10830$  transition is also important because it is particularly useful in constraining the
helium abundance, as shown e.g. by \cite{izotov_etal:2014} and subsequent authors. The emissivities for $\lambda 10830$ are summarised in Table~\ref{tab:table_10830} at three densities and three temperatures. The fuller tables of emissivities for temperatures of 5000K and 20000K are in the Appendix. The contribution of collisional excitation from the metastable $2~^3$S  depends directly on the population in that state. At low densities it is populated primarily by recombination followed by radiative decay to the ground state and the population increases with increasing density but as the density increases collisional de-excitation to the ground state and collisional excitation to higher singlet states cause the population to plateau once collisional de-excitataion dominates over radiative decay. The rising population of the metastable is reflected in the increasing emissivity of $\lambda 10830$ with increasing density and temperature.   The calculations of S96 and B99 used the effective collision strength data from \cite{sawey_berrington:1993} for this excitation process while P05, P12 and the present work took effective collision strengths from \cite{bray_etal:2000}. To isolate the effect of only changing the choice of effective collision strength, we can compare the results of our PB40 model with a model using the \cite{sawey_berrington:1993} data (SB93) at 10000K and at densities were collisional excitation of $\lambda 10830$ is dominant. The SB93 emissivities are larger by 4.3\%  at density $10^4$~cm$^{-3}$ and 4.9\% at density $10^6$~cm$^{-3}$. \cite{sawey_berrington:1993} and \cite{bray_etal:2000} tabulate their effective collision strengths on different temperature grids which only coincide  at 10000K, where we find that the \cite{sawey_berrington:1993} effective collision strengths for the $2~^3$P - $2~^3$S excitation  is larger than that of \cite{bray_etal:2000} by 4.2\% showing that changes to the collision strengths result in commensurate changes to the emissivity, as one might expect. \cite{benjamin_etal:1999} used the  \cite{sawey_berrington:1993} CE rates and their results agree better with our SB93 model than with PB40. 

There are, however, significant differences between our final results and those of P12, even though we used the same CE rates of \cite{bray_etal:2000}. As noted above, agreement is very good at the lowest density where radiative processes dominate and CE is not significant but not at the higher densities for some temperatures. The exception is the temperature of $10^4$K where the P12 emissivity is 1.6\% lower than ours at $10^4$~cm$^{-3}$ and agrees within 0.4\% at $10^6$~cm$^{-3}$. But at $5000$K the differences are 7.7\% and 13.4\% at those densities, while at 20000K the differences are -11.7\% and -16.3\%. \cite{bray_etal:2000} tabulate effective collision strengths on a logarithmic, base ten, mesh with interval  0.25 with the result that the temperature of 10000K is the only one of the three temperatures under discussion which corresponds to a tabulated value in their paper. The observation that P12 agrees well with the present results only at 10000K suggests that P12 have used a different approach to the interpolation of the tables of   \cite{bray_etal:2000} or have not used their results for all temperatures.

The features that we have seen when comparing our results to those of P12 for the 10830~\AA\ transition are present in most transitions in the emissivity tables at 5000, 10000 and 20000~K. Agreement is good at the lowest density with average absolute differences of 1.3\% or less. With increasing density the differences become larger, reaching +5.5\% at 5000~K and 16.1\% at 20000~K at density 10$^6$~cm$^{-3}$. These average differences conceal some very large individual differences between our work and P12. For example at 20000~K and a density of 10$^6$~cm$^{-3}$, the P12 values for the singlets are all larger than ours by a maximum of 66\%, this value being for $\lambda 6678$, and all smaller than ours by as much as 16.3\% for the triplets.  A similar trend but of smaller magnitude is seen at 10$^4$~cm$^{-3}$. Given that collisional excitation from the metastable is increasingly important for higher densities and temperatures, and despite the fact that P12 used \cite{bray_etal:2000} for collisional excitation, as did we, these differences are likely to be attributable to unidentified differences in the way that we and P12 calculated the rates for collisional processes among the lower levels.

The emissivities of the earlier P05 model are similar to those of P12,
except for the 5876, 6678~\AA\ lines, which were the most affected by the
code error previously mentioned
{(see also \citealt{aver_etal:2013}).}

We mentioned earlier the reasons for preferring the rates of \cite{bray_etal:2000}
over those of \cite{sawey_berrington:1993} and the fact that the \cite{bray_etal:2000} results are also not without weaknesses. In view of this, we consider that it is useful to view the differences between our two n=40 models, PB40 and SB93, that differ only in that they use these different CE rates, as a measure of the uncertainty due to this choice of rate coefficients, and as a way to identify lines that are relatively insensitive to that choice. For example Table~\ref{tab:table1} shows that the results of PB40 and SB93 for the strongest lines all agree within 1\% at 10$^2$~cm$^{-3}$, whereas in Table~\ref{tab:table2} at 10$^4$~cm$^{-3}$, eight of the lines, mostly but not exclusively among triplet states, differ by more than 1\%

The final column in the emissivity tables (PB+PE+PI) shows the result of adding photoionization and photoexcitation of excited states by a 100,000~K diluted black-body radiation field, in case B. With a dilution factor, d=10$^{-16}$, we can see that
 the emissivities of the main lines are very close to those of the PB model.
 We note that an appropriate dilution factor 0.1~pc from a typical white dwarf would be $\approx 5\times10^{-17}$ at which level PE and PI have a negligible effect on the emissivities.

The results of our Case B calculation, PB, are available from the CDS on a grid of electron temperatures and densities, log$_{10}$ N$_{\rm e}$[cm$^{-3}$]=2.0(0.5)6.0 and log$_{10}$ T$_{\rm e}$[K]=3.0(0.1)4.6, for all transitions with upper principal quantum number = 2-25 and lower principal quantum number = 2-5. We also provide a small FORTRAN program to make two-dimensional interpolation of the emissivity tables to any desired density and temperature within the tabulated ranges.


\subsection{Infrared lines}

\begin{table}
\begin{center}
  \caption{Emissivities of a selection of infrared lines, as listed in
    \citet{rubin_etal:1998}, and as calculated by S96 (case B) and with the
    present case B (PB). The wavelengths $\lambda$ are in vacuum except for 4471~\AA\, which is in air. 
    The values are relative to the  4471~\AA\ emissivity
  and were calculated for $T_{\rm e}$=10000 K and  $N_{\rm e}$=10000 cm$^{-3}$.
  The cases where the PB values differ from the S96 ones by more than 1\% are indicated in brackets.
  The last column (PBf) indicates our case B values, where the l-changing
  collision rates have been changed (see text). Changes of more than 1\% relative to PB are indicated.
  The last row gives the fluxes of the  4471~\AA\ line in 10$^{-26}$ erg cm$^{3}$ s$^{-1}$.
  \label{tab:rubin}    }
  \begin{tabular}{@{}lllll@{}}
  \hline 
$\lambda$  & levels    & S96  & PB    & PBf \\    
(microns)&     &      &     &      \\
  \hline
2.855020  & T 5p – 4s  & 0.00171  & 0.00171 &   0.00171 \\ 
3.330851  & S 5p – 4s  & 9.264 10$^{-4}$ & 9.32 10$^{-4}$  & 9.30 10$^{-4}$ \\
3.703571  & T 5d – 4p  & 0.005595 & 0.00563 &     0.00562 \\
4.006400 & S 5p – 4d  & 4.287 10$^{-4}$ & 4.32 10$^{-4}$ & 4.30 10$^{-4}$ \\
4.037735 & T 5f – 4d  & 0.02739  & 0.0268 & 0.0270 \\ %
4.040934 & S 5f – 4d  & 0.00913  & 0.00877 (4\%) & 0.00882 \\ 
4.049014 & T 5g – 4f  & 0.07561 & 0.0737 (3\%)   & 0.0761 (3\%) \\ 
4.049034 & S 5g – 4f  & 0.02520 & 0.0245  (3\%)   & 0.0253 (3\%) \\ %

4.054506 & S 5d – 4f  & 7.48 10$^{-5}$ & 4.90 10$^{-5}$ (53\%) & 4.89 10$^{-5}$ \\ 
4.056346 & T 5d – 4f  & 2.01 10$^{-4}$ & 1.84 10$^{-4}$ (9\%) & 1.84 10$^{-4}$ \\
4.122730  & S 5d – 4p  &  0.00222  & 0.00221 &  0.00221\\
4.244067  & T 5p – 4d  & 0.00312  &  0.00312  & 0.00311 \\
4.606601  & S 5s – 4p  & 7.479 10$^{-4}$ & 7.83 10$^{-4}$ (5\%) &  7.85 10$^{-4}$ \\
4.694980  & T 5s – 4p  & 0.001581 & 0.00187 (16\%) & 0.00188 \\
\\
0.4471  & T 4d-2p &  6.16  & 6.44 (4\%) & 6.43 \\
\hline
\end{tabular}
\end{center}
\end{table}

It is also interesting to compare our PB rates with those calculated by S96 for 
 weaker infrared lines, some of which were observed by \citet{rubin_etal:1998}.
 Table~\ref{tab:rubin} shows such a comparison, for the lines listed 
 by Rubin et al. The authors provide the S96 emissivities (case B) relative to 
 the 4471~\AA\ reference line, calculated for $T$=10000 K and  $N$=10000 cm$^{-3}$. The emissivity of the reference line in our PB 
 case is about 4\% higher than the S96 value. 
 Regarding the infrared lines, in most cases the relative intensities
 are within 1\% our values. However, there are a few notable deviations, particularly the 5d-4f transitions.

As we have previously mentioned, the rates for the electron and proton induced $l$-changing collisions
used by P05 and P12 are rather uncertain in that their calculations are based on CLOUDY but \cite{guzman_etal:2017} present proton rates that are stated to come from CLOUDY, which are too small by a factor of approximately forty. It is not clear whether these rates were used in the P05 and P12 calculations. It is clear, however, that those calculations did use the semi-classical methods of VF01 which grossly underestimate collision rates due to neglect of the quantum mechanical contribution from large impact parameters.  

As already shown by \cite{guzman_etal:2017}, the emissivities of the main optical
lines change by less than 0.1\%, when different electron rates are used,
with the exception of the case for a very low $T=100$ K and a high density 
of 10$^6$ cm$^{-3}$.  The same authors showed that variations of about 5\% 
are present in several cases for weaker transitions. 

To assess how much $l$-changing collisions affect the infrared lines in the Table, 
we have run our case B model decreasing the electron collision rates by a 
factor of six and also by decreasing  the proton ones by a factor of forty, consistent with the values in Figure~1 of  \cite{guzman_etal:2017}. 
The results, shown in the last column of Table~\ref{tab:rubin} (PBf),
indicate  
However, transitions from higher principal quantum numbers
are significantly affected. The effect of dramatically reducing the proton collision rates is to reduce the coupling between the populations of the higher-$l$ states and the lower-$l$ states, for a given principal quantum number. Since the low-$l$ states have the largest radiative decay rates this leads to larger departure coefficients in the reduced rate case for the higher-$l$ states and smaller departure coefficients for the lower-$l$ states. These effects only appear at intermediate principal quantum number, once the $l$-changing collision rates begin to dominate over radiative rates. As principal quantum number increases, these effects appear first in the high-$l$ states. For example, for $N_{\rm e}$ = 10000~cm$^{-3}$ and $T_{\rm e}$ = 10000K and when $l=n-1$ the reduced rates lead to an 8\% population increase at $n=10$ and an 9\% increase at $n=20$, while for $l=0$ decreases of a similar magnitude only appear for $n=40$ and greater. The  increases in the high-$l$ state populations are already apparent for $n=5$ in the 5g-4f transitions in Table~\ref{tab:rubin}.

\subsection{The Orion nebula model}

\begin{table*}
\begin{center}
   \caption{Emissivities of a selection from the strongest He optical lines. 
     The values are emissiviy ratios relative to   the 4471~\AA\ line.
     The third column gives the FOS-1SW, STIS-SLIT1c observed values, corrected for
     reddening; the fourth and fifth columns give the results of the CLOUDY model M
    and the case B predictions from \citet{porter_etal:2005}, as reported by
    \citet{blagrave_etal:2007}. Column PB gives our 
     case B solution, while the following ones give the results of the
     full model with PI and PE, for different dilutions d.
     All model emissivities in columns six, seven and eight have been calculated for
     $T_{\rm e}$=8000 K and  $N_{\rm e}$=2500 cm$^{-3}$.  Column nine (MD09) lists VLT observed intensities from
     \citet{mesa-delgado_etal:2009} and column ten (PB) our Case B results calculated for $T_{\rm e}$=8180 K and $N_{\rm e}$=2890 cm$^{-3}$. 
     Columns eleven (MD21) and twelve (PB) contain correspondng results from the VLT observations of \citet{mendez-delgado_etal:2021} calculated for $T_{\rm e}$=8360 K and 
     $N_{\rm e}$=5650 cm$^{-3}$. 
     For our calculations,
     we provide the emissivities of the 4471~\AA\ line in 10$^{-26}$ erg cm$^{3}$ s$^{-1}$
     in brackets. 
     \label{tab:table_orion}}
\begin{tabular}{@{}rllllllllllllllllllll@{}}
  \hline
$\lambda$ (\AA) & levels
                  &STIS-SLIT1c  & CLOUDY &P05 & PB &  d=10$^{-12}$ & d=10$^{-14}$ & MD09 & PB & MD21 & PB & \\
  \hline
 
2945 & T 5p-2s &  0.26$\pm$0.01   & 0.290 & 0.414 & 0.412 & 0.605 & 0.415 & - & 0.415 & - & 0.418 & \\
3188 & T 4p-2s &   -              & 0.441 & 0.878 & 0.862 & 1.197 & 0.866 & 0.802 & 0.875 & 0.471 & 0.886 & \\
3355 & S 7p-2s &   -              &       &       &       &       &       & 0.034 & 0.038 & 0.056 & 0.038 & \\
3448 & S 6p-2s &   -              &       &       &       &       &       & 0.056 & 0.061 & 0.079 & 0.061 & \\
3554 & T 10d-2p &  -              &       &       &       &       &       & 0.052 & 0.054 & 0.061 & 0.054 & \\
3587 & T 9d-2p &   -              &       &       &       &       &       & 0.077 & 0.075 & 0.081 & 0.075 & \\
3614 & S 5p-2s &   -              &       &       &       &       &       & 0.101 & 0.108 & 0.089 & 0.108 & \\
3634 & T 8d-2p &   -              &       &       &       &       &       & 0.099 & 0.108 & 0.106 & 0.107 & \\
3705 & T 7d-2p &   -              &       &       &       &       &       & 0.150 & 0.162 & 0.159 & 0.162 & \\
3820 & T 6d-2p &   -              &       &       &       &       &       & 0.262 & 0.262 & 0.251 & 0.262 & \\

3889 & T 3p-2s &   -              & 0.780 & 2.315 & 2.24  & 2.778 & 2.238 & 1.561 & 2.29 & 1.150 & 2.36 & \\ 
3965 & S 4p-2s &   -              & 0.206 & 0.221 & 0.219 & 0.214 & 0.219 & 0.211 & 0.221 & 0.194 & 0.222 & \\ 

4026 & T 5d-2p &   -              & 0.466 & 0.472 & 0.470 & 0.471 & 0.470 & - & 0.471 & 0.462 & 0.471 & \\
4144 & S 6d-2p &   -              &       &       &       &       &       & 0.065 & 0.069 & 0.070 & 0.069 & \\
4388 & S 5d-2p & 0.105$\pm$0.003  & 0.120 & 0.125 & 0.124 & 0.122 & 0.125 & 0.122 & 0124 & 0.119 & 0.124 & \\ 
4471 & T 4d-2p &   1              &  1    &  1    & (7.64)& (7.81)& (7.64) & 1 & (7.38) & 1 & (7.48) \\ 

4713 & T 4s-2p &   -              & 0.153 & 0.103 & 0.102 & 0.115 & 0.102 & 0.133 & 0.102 & 0.142 & 0.107 & \\ 
4922 & S 4d-2p &  0.204$\pm$0.005 & 0.258 & 0.272 & 0.269 & 0.264 & 0.270 & 0.278 & 0.269 & 0.262 & 0.268 & \\ 

5016 & S 3p-2s &  0.47$\pm$0.01   & 0.517 & 0.566 & 0.558 & 0.544 & 0.558 & 0.548 & 0.563 & 0.495 & 0.567 & \\ 
5876 & T 3d-2p &  2.90$\pm$0.04   & 2.815 & 2.789 & 2.88  & 2.842 & 2.880 & 3.014 & 2.87 & 3.033 & 2.89 & \\ 
6678 & S 3d-2p &  0.74$\pm$0.01   & 0.737 & 0.784 & 0.811 & 0.791 & 0.811 & 0.805 & 0.804 & 0.761 & 0.802 & \\ 

7065 & T 3s-2p &  1.46$\pm$0.02   & 1.714 & 0.612 & 0.560 & 0.618 & 0.560 & 1.247 & 0.593 & 1.527 & 0.671 & \\ 
7281 & S 3s-2p & 0.148$\pm$0.002   & 0.155 & 0.151 & 0.146 & 0.140 & 0.147 & 0.138 & 0.151 & 0.136 & 0.159 & \\
10028 & T 7f-3d & -               &       &       &       &       &       & 0.046 & 0.048 & 0.050 & 0.048 & \\
\hline
\end{tabular}
\end{center}
\end{table*}

{

The case B emissivities usually used to measure the  helium abundance are
susceptible to opacity effects, as e.g. discussed by \cite{robbins:1968,porter_etal:2007,blagrave_etal:2007}.
To illustrate this issue, we consider as an example the Orion nebula. 

In real nebulae, the stellar incident spectrum  would produce PI and PE
within  all the levels in neutral He. The neutral helium recombination
spectrum is obtained by a balance between  He$^{0}$, He$^{+}$, and He$^{++}$.
To approximate the case of a real nebula, and see which spectral
lines are more sensitive to the parameters of a real model, 
we have therefore considered PE within all levels and  PI 
from all levels up to $n=25$, although the results are the same if only PI 
from the ground state and the metastable are considered.

For case B, we have set the photo-excitation and de-excitations to the
ground state to zero. We do not model the effects of optical depth in lines or continuum.

We have adopted  a modelled photo-ionising spectrum obtained from the grid of O-star atmospheres by \cite{LanzHubeny:2003}
to represent the dominating  ionizing flux, from $\Theta$$^1$ Ori C,
an O6 star a radius about 9.4 $R_\Sun$. We use the model with an effective temperature of 40000 K and log $g$=4.0.

We  also experimented with a widely-used line-blanketed LTE atmosphere
 model spectrum from Kurucz, calculated with solar abundances, and surface
 gravity log g=4.5. The results are nearly the same, as
 this spectrum is very similar to the above, although with much lower spectral resolution. We note that in the 
 visible and infrared the spectra are close to that of a black-body of $T$=40,000 K,
 but the He ground state photoionizing flux below 504~\AA\ is very different.
We also note that the photons below the 2600~\AA\ threshold for photoionizing the metastable 2s $^3$S 
are important in driving the population of this level, which in turns 
affects the populations of all the triplets \citep[see, e.g.][]{clegg_harrington:1989}.
We have extended the input spectrum above 90 microns with that of a black-body of $T$=40,000 K.

\cite{blagrave_etal:2007} report an HST STIS 
observation (STIS-SLIT1C), for which they derive an approximate  temperature and 
density of  $T_{\rm e}$=8000 K and  $N_{\rm e}$=2500 cm$^{-3}$.
Table~\ref{tab:table_orion} presents in the first column the observed fluxes
corrected for reddening and relative to the 4471~\AA\ line.
 The second and third columns give the results of the CLOUDY model M
     and the case B (forced model) predictions 
     from \cite{porter_etal:2005} (P05), as reported by
     \cite{blagrave_etal:2007}.
As pointed out by the above-mentioned previous authors, 
opacity effects related to the population of the metastable 2s $^3$S are present.
In fact, the decays to this state such as the 3889 and 3188 are significantly 
over-predicted by the case B approximation (compared to observations),
indicating self-absorption. As a consequence, the decays to the 
2p $^3$P (as the 7065 and 4713~\AA\ lines)  are strongly under-predicted.
As shown by \cite{blagrave_etal:2007}, the CLOUDY model,
which takes into account (in a simplified way) opacity effects, provides
in general emissivities closer to observation.
Similar effects (although much reduced in size) are also present in 
the singlets. 

Returning to our model, we  list in column six of Table~\ref{tab:table_orion} the 
emissivities calculated with case B, column (PB).
As we have seen in previous tables, there is a general agreement
with the P05 results, although not at the level that one might expect
(we note that the 5876 and 6678~\AA\ lines were 
the most affected by code errors). 

We have then run our Case B model with PI and PE and various distances from the ionizing source.
\cite{baldwin_etal:1991} estimated that various regions of the Orion nebula
range in distance from $\Theta ^1$ Ori C between 3 $\times 10^{17}$ and  
$10^{18}$ cm, which result in dilution factors of about 
$2.5\times10^{-12}$ and  $2.2\times 10^{-13}$. We show in Table~\ref{tab:table_orion}
the results of the full model at representative dilutions of $10^{-12}$ and $10^{-14}$. 
The Table clearly indicates which lines are most affected by the varying 
PI and PE processes. It is worth noting that variations of a 
few percent are also seen in the singlets. 
Also, the values in the Table 
show  that for sufficiently large distances we recover virtually the same emissivities as
obtained with the standard Case B calculation.

\cite{mesa-delgado_etal:2009} and \cite{mendez-delgado_etal:2021} report high-resolution deep spectroscopic VLT observations of Herbig-Haro objects in the Orion Nebula which also include spectra of the nebula itself. These spectra contain many helium lines beyond the usual strong visible transitions, some from relatively high principal quantum number. For extraction,  \cite{mendez-delgado_etal:2021} divide their slit into four "cuts", with cut~4 exclusively containing nebular material. Dereddened fluxes for some of the stronger helium lines are listed under MD09 and MD21 in Table~\ref{tab:table_orion}, accompanied by our Case B results calculated at the temperatures and densities derived for the nebular material by the respective authors. For those transitions which are expected to have a negligible contribution from CE or optical depth effects related to the metastable, agreement is generally good or very good especially for  MD09. For example, the MD09 intensities for transitions within  the singlet states  agree with theory within the stated observational errors for all transitions listed in  Table~\ref{tab:table_orion}. The same applies to the 7f-3d and $n$d-2p triplet transitions for MD09. It is, however, not true for the intensities from MD21, where significant differences are present in some cases. For example, the intensities of the singlet $n$p-2s series fall well below theory and outside the error bars for $n=3,4$ and 5 but significantly above theory for $n=6$ and 7. The $n$d-2p intensities are also larger than theory and outside the stated error bars for $n$= 9 and 10. However, the very good agreement between theory and the results of MD09 give confidence that the theory is reliable for those transitions where the effects of population in the metastable are not significant. The reason for the less good agreement with theory for MD21 is unclear but given the good agreement with MD09, it seems unlikely to be due to errors in theory. Observed intensities in the singlet $n$p-2s series being below Case B predictions could be explained by some $n$p-1s photon escape leading to results tending towards Case A but this effect should get stronger as $n$ increases, while the opposite is observed. 

As mentioned above, self-absorption in transitions ending on the 2$^3$S metastable weakens the triplet $n$p-2s series. This makes it difficult to compare observation and theory for individual triplet lines, especially the strongest visible lines. The effect is strongest for the 3p-2s $\lambda 3889$ transition and leads to an increase in the 3s-2p $\lambda 7065$ intensity. If we neglect the effect of opacity on the states with $n>3$ we would expect the combined energy in these two transitions to be the same irrespective of optical depth effects. From Table~\ref{tab:table_orion} we find the observed $I(3889)+I(7065)$ to be 2.81 and 2.88 in Case B from the MD09 observations. The corresponding numbers for the MD21 data are 2.68 and 3.03. In the MD09 data the 4p-2s $\lambda 3188$ intensity is lower than the Case B prediction by ~8\%. If this is due to self-absorption we would expect enhancements in 3d-2p $\lambda 5876$ and also 3s-2p $\lambda 7065$. Adding the intensities of all five lines we find an observed value of 6.75 from the MD09 observations and 6.73 from our Case B calculations, which is excellent agreement.  The corresponding results for the MD21 observations are 6.32 and 6.91 but the large difference between observation and theory for the 4p-2s $\lambda 3188$ transition indicates that this transition is showing strong self-absorption and even higher members of $n$p-2s series would need to be included to make a valid comparison with theory.

}

\section{Conclusions}

{ 

Having established in a previous work \citep{delzanna_etal:2020_he} that earlier 
studies of the He recombination spectrum suffered from various shortcomings,
we have built several collisional-radiative models and compared the 
emissivities of the main spectral lines in neutral He 
with some of the most widely-used values in the literature.
We have focused the comparisons on the case B approximation, as in most previous work. 

As the requirement on the accuracy of the predicted emissivities is
stringent, of the order of 1\%, we conclude that there are several
problems in the previous studies. 
A detailed assessment on the reasons why significant (larger than 1\%)
differences among different models are present is difficult.
The adoption of different CE rates clearly has an effect,
but the differences between the S96 and B99 models are surprising.

Before the various implementations of different models in CLOUDY,
the S96 was the largest recombination model for He.
Comparing the results of a reduced model ($LS$-resolved states up to $n$=40) 
to a full model, we were able to partially validate the S96 assumption,
showing that for most spectral lines {within the low-lying states the}  emissivities are accurate
(within 1\%) with the reduced model. So for most cases, a larger model
is not strictly necessary.
Despite this, significant differences with the S96 results are found,
especially for a selection of infrared lines.

The latest implementation within CLOUDY, described in P12,
used more accurate rates than previous versions, but still
included incorrect rates for $l$-changing collisions.  Using our models we found that 
in reality  such rates have little
effect on the emissivities of the main optical transitions.
{A similar conclusion was found by \cite{guzman_etal:2017} when varying the 
electron collision rates}.
There is generally excellent agreement between our emissivities and
the P12 ones at the lowest densities, irrespective of temperature, where radiative processes dominate the populations but significant discrepancies are present
for several transitions for higher densities at all temperatures.

The principal remaining uncertainty in our atomic model is now the choice of CE rate coefficients from the 2~$^3$S metastable, which becomes increasingly significant as density increases. Of the two most accurate calculations of these coefficients by \cite{sawey_berrington:1993} and \cite{bray_etal:2000} we have chosen to use the latter but with the caveat that both have weaknesses. We suggest that it is reasonable to view the differences in emissivities resulting from using one or other of these CE calculations as a measure of the uncertainty due to the CE rates. In particular, by comparing our results from the SB93 model with our PB40 model it is possible to identify transitions which are relatively insensitive to the choice of CE rates. Broadly, the relatively weaker intersystem CE rates means that transitions among singlet states are less sensitive than those within the triplets, although there are some CE effects even on the singlets for transitions with an $n$=3 upper state. Within the triplet states themselves, transitions are preferred where the upper state is not linked to the metastable by an an electric dipole transition, such as the $n$d-2p series. The 4f-3d and 5g-4f transitions in the IR are another such example where CE effects are expected to be very small. At the highest density we consider, the differences between emissivities from the SB93 and PB40 models reach 5\% for the $\lambda 10830$ transition, which is the one most affected by CE.

Aside from the above issues, there is a further major one which has often 
been overlooked in the literature. The tabulated He emissivities calculated with the un-physical case B forced solution are
widely used in astrophysical codes \citep[see,e.g.][]{luridiana_etal:2015}
and in general within the literature to e.g. measure the helium abundance 
\citep[see,e.g.][for some recent examples]{peimbert_etal:2007,aver_etal:2013,izotov_etal:2014,aver_etal:2015,peimbert_etal:2016}. 

A full analysis should include observed or well-modeled photoionizing 
radiation fields, updated atomic data, all the PI and PE effects we have included, 
and solve the full radiative transfer problem including the 
any nebular expansion, which have a significant effect as shown by 
\cite{robbins:1968}. In some cases, where optical depths are not too great, the sum of the intensities of a subset of the triplet lines can be a useful diagnostic.

However, for those cases where opacity and other effects are 
negligible, and the case B solution is an acceptable approximation
we provide our results in electronic form, including all transitions within $n \le 5$ 
and all those between the $n \le 5$ and $n' \le 25$ states.  
{We also provide an interpolation program. }
We note that we have found significant 
differences between our emissivities and those calculated by 
S96 for a few infrared transitions, discussed by \cite{rubin_etal:1998}.
We therefore recommend our emissivities for future studies, and 
to benchmark any new case B model.

}

\section*{Acknowledgments}
GDZ acknowledges support from STFC (UK) via the consolidated grants 
to the atomic astrophysics group (AAG) at DAMTP, University of Cambridge (ST/P000665/1. and ST/T000481/1).
{We would like to thank N.R. Badnell for useful discussions on the 
calculations of A-values and rates for l-changing electron collisions and the referee for some helpful comments.
}

\section*{Data Availability}
The results of our Case B calculation, PB, are available from the CDS on a grid of electron temperatures and densities, log$_{10}$ N$_{\rm e}$[cm$^{-3}$]=2.0(0.5)6.0 and log$_{10}$ T$_{\rm e}$[K]=3.0(0.1)4.6, for all transitions with upper principal quantum number = 2-25 and lower principal quantum number = 2-5. We also provide a  FORTRAN program to make two-dimensional interpolation of the emissivity tables.

\bibliographystyle{mn2e}

\bibliography{paper}


\appendix

\section{Other  cases}

 \setlength{\tabcolsep}{2pt}

\begin{table*}
\begin{center}
  \caption{Emissivities  (10$^{-26}$ erg cm$^{3}$ s$^{-1}$ ) of the
            strongest He lines, for $T_{\rm e}$=20000 K and  $N_{\rm e}$=10$^{6}$ cm$^{-3}$.
\label{tab:table1}
  }   
   \begin{threeparttable}
\begin{tabular}{rllllllllllll}
  \hline
 $\lambda$ (\AA) & levels & \multicolumn{7}{c}{Earlier work} & \multicolumn{4}{c}{Present work} \\ 
 \cmidrule(lr){3-9}\cmidrule(lr){10-13}
        &       & B72 (A)&B72 (B)&S96 (A)&S96 (B)& B99 (B) &P05 (B)& P12 (B) & SB93(B) & PB40 & PB (A) & PB (B) \\
    \hline
2945 & T 5p-2s & 1.66   & 1.66  &  -    & -      & 1.87   & 2.11   & 2.22   & 2.26 & 2.27 & 2.27 & 2.27   \\
3188 & T 4p-2s & 3.43   & 3.43  & 3.47  & 3.47   & 5.47   & 4.96   & 5.09  & 5.87 & 5.08 & 5.08 & 5.09 \\

3889 & T 3p-2s & 8.30   & 8.30  & 8.22  & 8.22   & 16.7   & 14.9   & 15.0   & 17.8 & 16.2 & 16.2 & 16.3   \\
3965 & S 4p-2s & 0.027  & 0.83  &       & 0.81   & 0.99   & 1.49   & 1.21   & 1.07 & 1.02 & 0.034& 1.02    \\

4026 & T 5d-2p & 1.45   & 1.45  & 1.44  & 1.44   & 1.44   &  1.90  & 2.01   & 2.04 & 2.05 & 2.05  & 2.05  \\
4388 & S 5d-2p & 0.37   & 0.38  &       & 0.38   & 0.38   &  0.47  & 0.57   & 0.45 & 0.45 & 0.43  & 0.45  \\
4471 & T 4d-2p & 2.98   & 2.98  & 3.0   & 3.0    & 5.58   &  4.42  & 4.71   & 6.01 & 4.72 & 4.72  & 4.72  \\

4713 & T 4s-2p & 0.41   & 0.41  & 0.46  & 0.46   & 1.72   &  1.09  & 1.05  & 1.77 & 1.31 & 1.31  & 1.31  \\
4922 & S 4d-2p & 0.78   & 0.80  & 0.76  & 0.79   & 1.05   &  1.03  & 1.31   & 1.10 & 0.97 & 0.93  & 0.97  \\

5016 & S 3p-2s & 0.045  & 2.04  & 0.045 & 1.99   & 2.71   &  2.74  & 3.17   & 2.87 & 2.76 & 0.062  & 2.76 \\
5876 & T 3d-2p & 7.62   & 7.62  & 7.56  & 7.56   & 17.8   &  16.6  & 18.3   & 18.9 & 17.6& 17.6  & 17.6   \\
6678 & S 3d-2p & 2.09   & 2.14  & 2.05  & 2.12   & 2.99   &  3.25  & 4.88   & 3.10 & 2.93& 2.82  & 2.93  \\

7065 & T 3s-2p & 1.40   & 1.40  & 1.91  & 1.91   & 9.15   &  8.10  & 7.62   & 9.40 & 8.85& 8.83  & 8.85 \\
7281 & S 3s-2p & 0.33   & 0.55  & 0.34  & 0.58   & 1.36   &  1.23  & 1.30   & 1.40 & 1.30& 0.99  & 1.30  \\
10830 & T 2p-2s & 14.9  & 14.9  & 351   & 351    & 253    & 237    & 215    & 257 & 256 & 256  & 256 \\
18685 & T 4f-3d &       &       &       &        & 1.28   & 1.32  & 1.78   & 1.36 & 1.25& 1.24  & 1.25  \\
20587 & S 2p-2s &
2.6$\times$10$^{-4}$     &       &       & 6.28   & 5.46   &   -    & 5.98   & 5.33 &  5.6& 6$\times$10$^{-3}$& 5.55 \\
\noalign{\smallskip}
\hline
\end{tabular}
\begin{tablenotes}
    \item[] {The first column gives the wavelength
  (in air, except the  last ones in vacuum), the second indicates if a line is
between singlets (S) or triplets (T).
B72 (A): Brocklehurst (1972) case A;
   B72 (B): Brocklehurst (1972)  case B    ;
      S99 (A): Smits (1999), case A; S99 (B): Smits (1999), case B; 
      B99(B): Benjamin et al. (1999) case B;         
         P05 (B): Porter et al. (2005) case B;
         P12 (B): Porter et al. (2012) case B;
         SB93(B): n=40 model with SB93 rates, case B;
         PB40 (B): n=40 model with the Bray et al. rates, case B;
         PB(A): full n=500 model case A;  PB(B): full n=500 model case B.
}
 \end{tablenotes}
\end{threeparttable}
\end{center}
\end{table*}

\begin{table}
\begin{center}
   \caption{Emissivities  (10$^{-26}$ erg cm$^{3}$ s$^{-1}$) of the
     strongest He lines, for $T{\rm e}$=20000 K and  $N_{\rm e}$=10$^{4}$ cm$^{-3}$
     case B.
\label{tab:table2} }
\begin{tabular}{@{}rllllllllllllllllllll@{}}
  \hline
 $\lambda$ (\AA) & levels & \multicolumn{5}{c}{Earlier work} & \multicolumn{2}{c}{Present work} \\ 
 \cmidrule(lr){3-7}\cmidrule(lr){8-9}
  &   & B72  &S96 & B99    &P05    & P12 & PB40 & PB &  \\
\hline  
2945 & T 5p-2s & 1.65&   -   & 1.82   & 1.99   & 2.12 & 2.13 & 2.13 &  \\
3188 & T 4p-2s & 3.41& 3.45  & 5.05   & 4.58   & 4.79 & 4.71 & 4.71 &   \\

3889 & T 3p-2s & 8.22& 8.18  & 14.97   & 13.3  & 13.8 & 14.5 & 14.5 &  \\
3965 & S 4p-2s & 0.82& 0.81  & 0.95    & 0.97  & 1.15 & 0.97 & 0.97 &   \\

4026 & T 5d-2p & 1.43& 1.44  & 1.44   &  1.78  & 1.92 & 1.92 & 1.92 &   \\
4388 & S 5d-2p & 0.38& 0.38  & 0.38   &  0.45  & 0.55 & 0.43 & 0.43 &  \\
4471 & T 4d-2p & 2.95& 3.0   & 5.07   &  4.08  & 4.44 & 4.34 & 4.35 &   \\

4713 & T 4s-2p & 0.41& 0.46  & 1.48   &  0.96  & 0.95 & 1.13 & 1.13 &  \\
4922 & S 4d-2p & 0.80& 0.79  & 0.99   &  0.97  & 1.25  & 0.93 & 0.93 &  \\

5016 & S 3p-2s & 2.03& 1.98  & 2.55   &  2.55  & 2.99 & 2.58 & 2.58 &   \\
5876 & T 3d-2p & 7.60& 7.56  & 15.8   &  14.8  & 17.0 & 15.5 & 15.5 &  \\
6678 & S 3d-2p & 2.15& 2.14  & 2.81   &  3.08  & 4.84 & 2.74 & 2.74 &  \\

7065 & T 3s-2p & 1.40& 1.92  & 7.73   &  6.81  & 6.65 & 7.36 & 7.36 &  \\
7281 & S 3s-2p & 0.55& 0.58  & 1.20   &  1.09  & 1.18 & 1.14 & 1.14 &   \\
10830 & T 2p-2s& 14.8& 272.7 & 206.6  &  188.4 & 181  & 204  & 204 &  \\

18685 &T 4f-3d &  -&   -     & 1.21   &  1.26  & 1.78 & 1.17 & 1.17 &   \\
20587 &S 2p-2s &  -&  4.95   & 4.56   &   -    & 5.16 & 4.50 & 4.50 &   \\
\hline
\end{tabular}
\end{center}
\end{table}

\begin{table}
\begin{center}
   \caption{Emissivities  (10$^{-26}$ erg cm$^{3}$ s$^{-1}$ ) of the
     strongest He lines, for $T{\rm e}$=20000 K and  $N_{\rm e}$=10$^{2}$ cm$^{-3}$
     case B.
\label{tab:table3}
}
\begin{tabular}{@{}rllllllllllllllllllll@{}}
\hline 
$\lambda$ (\AA) & levels & \multicolumn{5}{c}{Earlier work} & \multicolumn{2}{c}{Present work} \\ 
 \cmidrule(lr){3-7}\cmidrule(lr){8-9}
            &    &  B72   &S96 & B99     &P05    & P12    & PB40  &  PB  &  \\
\hline
2945 & T 5p-2s  & 1.65& -     & 1.68   & 1.65   & 1.69   & 1.68 & 1.68 &  \\
3188 & T 4p-2s  & 3.40& 3.45  & 3.53   & 3.43   & 3.50   & 3.48 & 3.48 &   \\

3889 & T 3p-2s  & 8.20& 8.17  & 8.53   & 8.32   & 8.61   & 8.59 & 8.58 &   \\
3965 & S 4p-2s  &    -& 0.80  & 0.81   & 0.81   & 0.84   & 0.82 & 0.82 &   \\

4026 & T 5d-2p  & 1.43& 1.44  & 1.44   &  1.46  & 1.48   & 1.47 & 1.47 &  \\
4388 & S 5d-2p  &    -& 0.38  & 0.38   &  0.38  & 0.39   & 0.38 & 0.38 &   \\
4471 & T 4d-2p  & 2.94& 3.00  & 3.11   &  3.01  & 3.05   & 3.03 & 3.03 &  \\

4713 & T 4s-2p  & 0.41& 0.46  & 0.51   &  0.48  & 0.49   & 0.49 & 0.49 &  \\
4922 & S 4d-2p  &    -& 0.79  & 0.80   &  0.80  & 0.82   & 0.80 & 0.80 &  \\

5016 & S 3p-2s  &    -& 1.97  & 2.01   &  2.00  & 2.06   & 2.03 & 2.03 &   \\
5876 & T 3d-2p  & 7.58& 7.56  & 8.06   &  7.89  & 8.13   & 7.99 & 8.00 &   \\
6678 & S 3d-2p  &    -& 2.14  & 2.18   &  2.17  & 2.30   & 2.17 & 2.18 &   \\

7065 & T 3s-2p  & 1.40& 1.92  & 2.22   &  2.15  & 2.18   & 2.18 & 2.18 &   \\
7281 & S 3s-2p  &    -& 0.58  & 0.61   &  0.59  & 0.61   & 0.61 & 0.61 & \\
10830 & T 2p-2s &  14.7& 25.7 & 24.6   &  23.4  & 23.8   & 23.7 & 23.6 &  \\

18685 &T 4f-3d  &    -&  -    & 0.92   &  0.96  & 0.94   & 0.91 & 0.92 &  \\
20587 &S 2p-2s  &    -&  2.24 & 2.24   &   -    & 2.29   & 2.24 & 2.24 &   \\
\hline
\end{tabular}
\end{center}
\end{table}


\begin{table}
\begin{center}
   \caption{Emissivities  (10$^{-26}$ erg cm$^{3}$ s$^{-1}$ ) of the
     strongest He lines, for $T{\rm e}$=5000 K and  $N_{\rm e}$=10$^{6}$ cm$^{-3}$
     case B. %
\label{tab:table4}
}
\begin{tabular}{@{}rllllllllllllllllllll@{}}
\hline 
$\lambda$ (\AA) & levels & \multicolumn{5}{c}{Earlier work} & \multicolumn{2}{c}{Present work} \\ 
 \cmidrule(lr){3-7}\cmidrule(lr){8-9}
 &   & B72 & S96   & B99    &   P05  & P12 & PB40 & PB  & \\
\hline

2945 & T 5p-2s  & 4.30& -     & 4.36   & 4.57   &  4.57  & 4.36 & 4.36 &   \\
3188 & T 4p-2s  & 9.04& 9.10  & 9.18   & 9.59   &  9.55  & 9.12 & 9.12 &   \\

3889 & T 3p-2s  & 22.9& 22.7  & 23.3   & 25.0   &  24.6  & 23.6 & 23.6 &   \\
3965 & S 4p-2s  & 2.41& 2.36  & 0.95   &  2.53  &  2.53  & 2.41 & 2.41 &   \\

4026 & T 5d-2p  & 5.56& 5.45  & 5.47   &  5.87  &  5.78  & 5.57 & 5.56 &   \\
4388 & S 5d-2p  & 1.51& 1.45  & 1.46   &  1.57  &  1.56  & 1.49 & 1.48 &   \\
4471 & T 4d-2p  & 12.0& 12.0  & 12.1   & 12.70  &  12.5  & 12.0 & 12.0 &  \\

4713 & T 4s-2p  & 0.77& 0.90  & 1.48   &  0.98  &  0.95  & 0.94 & 0.93 &  \\
4922 & S 4d-2p  & 3.32& 3.23  & 3.25   &  3.46  &  3.43  & 3.26 & 3.26 &   \\

5016 & S 3p-2s  & 6.20& 6.03  & 6.11   &  6.50  &  6.49  & 6.15 & 6.16 &   \\
5876 & T 3d-2p  & 35.4& 34.8  & 35.3   &  36.9  &  37.4  & 35.1 & 35.0 &   \\
6678 & S 3d-2p  & 10.2& 10.0  & 10.1   &  10.5  &  10.7  & 10.0 & 10.0 &   \\

7065 & T 3s-2p  & 2.89& 4.29  & 5.19   &  5.30  &  5.22  & 5.05 & 5.04 &   \\
7281 & S 3s-2p  & 1.39& 1.37  & 1.46   &  1.50  &  1.48  & 1.43 & 1.43 &   \\
10830 & T 2p-2s & 47.5& 245.  & 234.   &  214   &  213   & 189  & 189 &   \\

18685 T& 4f-3d  &    -& -    & 4.97   &   5.41  &  5.35 &  4.90 & 4.90 &   \\
20587 S& 2p-2s  &    -&10.7  & 10.6   &   -     & 10.7  &  10.1 & 10.1 &   \\

\hline
\end{tabular}
\end{center}
\end{table}

\begin{table}
\begin{center}
   \caption{Emissivities  (10$^{-26}$ erg cm$^{3}$ s$^{-1}$ ) of the
     strongest He lines, for $T{\rm e}$=5000 K and  $N_{\rm e}$=10$^{4}$ cm$^{-3}$
     case B.
\label{tab:table5}
}
\begin{tabular}{@{}rllllllllllllllllllll@{}}
\hline 
$\lambda$ (\AA) & levels & \multicolumn{5}{c}{Earlier work} & \multicolumn{2}{c}{Present work} \\ 
 \cmidrule(lr){3-7}\cmidrule(lr){8-9}
 &                 &  B72 &   S96  & B99 & P05    & P12    & PB40 &  PB & \\
\hline
2945 & T 5p-2s  & 4.16&    -   & 4.21 & 4.26   & 4.26  & 4.22 & 4.22 &   \\
3188 & T 4p-2s  & 8.76& 8.82   & 8.83 & 8.95   & 8.92 & 8.83 & 8.82 &   \\

3889 & T 3p-2s  & 22.2& 22.0   & 22.3 & 23.3   & 22.9  & 22.7 &  22.7 &   \\
3965 & S 4p-2s  & 2.33& 2.29   & 2.29 &  2.35  & 2.36  & 2.32 &  2.32 &   \\

4026 & T 5d-2p  & 5.38& 5.32   & 5.31 &  5.43  &  5.43 & 5.41 &  5.41 &   \\
4388 & S 5d-2p  & 1.46& 1.41   & 1.42 &  1.45  &  1.46 & 1.44 &  1.44 &   \\

4471 & T 4d-2p  & 11.7& 11.8   & 11.8 & 11.8   &  11.8 & 11.7 & 11.7 &  \\

4713 & T 4s-2p  & 0.76& 0.88   & 0.90 &  0.93  &  0.90 & 0.91 &  0.91 &  \\
4922 & S 4d-2p  & 3.23&  3.09  & 3.17 &  3.22  &  3.21 & 3.19 &  3.19 &   \\

5016 & S 3p-2s  & 6.00&  0.13  & 5.88 &  6.04  & 6.04  & 5.96 &  5.96 &   \\
5876 & T 3d-2p  & 35.3& 34.8   & 35.0 &  34.4  & 34.9  & 35.2 &  35.2 &   \\
6678 & S 3d-2p  & 10.2& 9.8    & 10.0 &  9.9   & 10.0  & 10.1 &  10.1 &   \\

7065 & T 3s-2p  & 2.85& 4.20  & 4.70  &  4.75  & 4.67  & 4.67 &  4.66 &   \\
7281 & S 3s-2p  & 1.26& 0.64  & 1.39  &  1.39  & 1.38  & 1.38 &  1.39 &   \\
10830 & T 2p-2s & 46.4& 154   & 152   &  140   & 140   & 130  &  130  &   \\

18685 T& 4f-3d  &    -&   -   & 5.06  &   5.13 &  4.99 & 5.09 & 5.09 &   \\
20587 S& 2p-2s  &    -&8.96   & 8.94  &   -    &  8.85 & 8.82 & 8.83  &   \\
\hline
\end{tabular}
\end{center}
\end{table}

\begin{table}
\begin{center}
   \caption{Emissivities  (10$^{-26}$ erg cm$^{3}$ s$^{-1}$ ) of the
     strongest He lines, for $T{\rm e}$=5000 K and  $N_{\rm e}$=10$^{2}$ cm$^{-3}$.
     case B.
\label{tab:table6}
}
\begin{tabular}{@{}rllllllllllllllllllll@{}}
\hline 
$\lambda$ (\AA) & levels & \multicolumn{5}{c}{Earlier work} & \multicolumn{2}{c}{Present work} \\ 
 \cmidrule(lr){3-7}\cmidrule(lr){8-9}
&      &   B72  &   S96 & B99  & P05  & P12 & PB40 &  PB  & \\
\hline
2945 & T 5p-2s  & 4.12&    -   & 4.20 & 4.14   & 4.19  &  4.19 & 4.17 &   \\
3188 & T 4p-2s  & 8.66& 8.76   & 8.79 & 8.69   & 8.76  &  8.75 & 8.72 &   \\

3889 & T 3p-2s  & 21.9& 21.8   & 22.0 & 22.5   &  22.4  & 22.3 & 22.3 &   \\
3965 & S 4p-2s  &    -& 2.27   & 2.28 &  2.28  &  2.31  & 2.30 & 2.30 &   \\

4026 & T 5d-2p  & 5.31& 5.28   & 5.31 &  5.28  &  5.37  & 5.35 & 5.35 &   \\
4388 & S 5d-2p  &    -& 1.41   & 1.42 &  1.41  &  1.44  & 1.43 & 1.43 &   \\

4471 & T 4d-2p  & 11.6& 11.74  & 11.8 & 11.5   &  11.6  & 11.6 & 11.6 &  \\

4713 & T 4s-2p  & 0.76& 0.88   & 0.89 &  0.90  &  0.89  & 0.90 & 0.89 &  \\
4922 & S 4d-2p  &    -& 3.16   & 3.17 &  3.13  &  3.18  & 3.16 & 3.16 &   \\

5016 & S 3p-2s  &    -& 5.81   & 5.85  &  5.85  & 5.93  & 5.89 & 5.89 &   \\
5876 & T 3d-2p  & 35.0& 34.5   & 35.3  &  33.6  & 34.9  & 34.9 & 35.2 &   \\
6678 & S 3d-2p  &    -& 10.05  & 10.11 &  9.64  & 10.0  & 10.0 & 10.1 &   \\

7065 & T 3s-2p  & 2.83& 4.18   & 4.22  &  4.27  & 4.24  & 4.25 & 4.23 &   \\
7281 & S 3s-2p  &    -& 1.33   & 1.35  &  1.32  & 1.33  & 1.34 & 1.33 &   \\
10830 & T 2p-2s & 46.0& 50.6   & 50.7  &  49.9  & 50.8  & 50.6 & 50.6 &   \\

18685 T& 4f-3d  &    -&   -    & 5.19  &   5.06 & 5.07  & 5.10 & 5.19 &   \\
20587 S& 2p-2s  &    -& 7.42   & 7.42  &   -    & 7.42  & 7.45 & 7.47 &   \\

\hline
\end{tabular}
\end{center}
\end{table}

\end{document}